\documentclass[preprint,12pt]{elsarticle}
\usepackage{amssymb}
\usepackage{color}

\journal{Optics Communications}

\begin{document}

\begin{frontmatter}

\title{Prospect for detecting squeezed states of light created by a
  single atom in free space}

\author[de,pl]{Magdalena Stobi\'nska} \ead{magda.stobinska@mpl.mpg.de}
\author[de]{Markus Sondermann} \author[de]{Gerd Leuchs}

\address[de]{Institute for Optics, Information and
  Photonics, Erlangen-N\"urnberg University, Staudtstr. 7/B2, 91058
  Erlangen, Germany,\\ and Max Planck Institute for the Science of
  Light, G\"unter-Scharowsky-Str.\ 1, Bau 24, 91058 Erlangen, Germany}

\address[pl]{Institute of Theoretical Physics and Astrophysics, University
  of Gda\'nsk, Poland }

\begin{abstract}
We discuss the possibilities of studying in detail the dynamics of
spontaneous emission of a single photon by a single atom and
measuring the transient degree of squeezing by means of full solid
angle fluorescence detection.
\end{abstract}

\begin{keyword}
squeezing, resonance fluorescence, single photon, single atom
\end{keyword}

\end{frontmatter}

\section{Introduction}

\begin{quote}\it
  We are mourning our teacher and colleague Krzysztof
  W\'odkiewicz whom we will fondly remember as a skillful scientist,
  sharp thinker and friend.
\end{quote}

Quadrature squeezed states of light are a basic resource for
continuous variable (CV) based quantum communication.  Two independent
CV squeezed states interfering on a beam splitter form at its output
CV Einstein-Podolsky-Rosen entangled
states~\cite{Yeoman1993,Leuchs1999}. Experimentally this entanglement
was demonstrated for vacuum states~\cite{Ou1992} and for intense laser
beams \cite{Silberhorn2001}. Based on squeezed-state entanglement
quantum teleportation of coherent states was experimentally achieved
\cite{Furusawa1998}.  Recently it was shown that a single quadrature
squeezed photon also enables obtaining teleportation of coherent state
qubits \cite{Branczyk2008}. Moreover, quadrature squeezing can be
transformed into polarization squeezing \cite{Heersink2003}.  CV
squeezed states have also been useful for developing various other
tools for quantum information processing. For example, they have been
used for realizing quantum nondemolition coupling and thus CV
quantum erasing \cite{Andersen2004}. Furthermore, a method for dense
quantum coding for the quadrature amplitudes of the electromagnetic
field has been proposed \cite{Braunstein2000}.

Whereas in the majority of the references mentioned so far
squeezed states have been used as a tool, we want to focus on a more
fundamental aspect here.
We propose an experiment that is expected to
enable the successful observation of a squeezed state that is
generated in one of the most fundamental settings of quantum optics:
the spontaneous emission of a single photon by a single two-level
atom that is prepared in a suitable superposition state.

The name `squeezed' states emerges from their property that the quantum
uncertainty in one of two noncommuting observables $[A,B]=i C$ in these
states is decreased $(\Delta A)^2 < \frac{1}{2}|\langle C \rangle|$ at
the cost of increasing uncertainty in the other one $(\Delta B)^2 >
\frac{1}{2}|\langle C \rangle|$, in order to obey the Heisenberg
uncertainty relation $\Delta A \Delta B \ge \frac{1}{2}|\langle C
\rangle|$. Amplitude and phase quadrature operators are defined as 
\begin{equation}
X_1 = 1/2(a + a^{\dagger}), \quad X_2 = i/2(a^{\dagger}-a),
\end{equation}
where $a$ and $a^{\dagger}$ denote annihilation and creation operators
respectively, $X_{1,2}$ are noncommuting observables $[X_1,X_2]=i/2$
for which $\Delta X_1 \Delta X_2 \ge 1/4$. Therefore, a state for
which $(\Delta X_k)^2< 1/4$ for $k=1,2$ is called quadrature
squeezed. 

Usually quadrature squeezed states are produced via a squeezing
transformation generated by a Hamiltonian quadratic in annihilation
$a$ and creation $a^{\dagger}$ operators applied to a coherent state
$|\alpha\rangle$. For a single mode of light the transformation takes
the following form
\begin{equation} 
\hat{S} = \exp\left\{\frac{\xi^*}{2} a^2 - \frac{\xi}{2}\
  {a^{\dagger}}^2\right\}, 
\label{sq}
\end{equation}
where $\xi$ is called the squeezing parameter.  

Early on it was
realized that applying the operator $\hat{S}$ to the vacuum state
leads to a superposition of even number Fock states~\cite{Yuen1976},
hence the name two-photon coherent states for this special class of
squeezed states.  In experiments squeezing is realized either in the
parametric amplification or in four wave mixing process in
$\chi^{(3)}$ nonlinear medium (e.g. in optical fiber).

Squeezing is present not only in a superposition of two macroscopic coherent
states \cite{Schleich1991} but also in the superposition of vacuum
and a single photon arising in the process of spontaneous emission of
an atom into a single cavity mode \cite{Wodkiewicz1987}. 
At the time this came a bit as a
surprise\footnote{One of us (GL) vividly remembers Krzysztof
  W\'odkiewicz coming to his office at the Max Planck Institute for
  Quantum Optics in Garching emphasizing that he had just found an
  unexpected result: The superposition of the $|0\rangle$ and the
  $|1\rangle$ Fock states may show squeezing. The surprise was that
  squeezed states generated from vacuum were not exclusively
  superpositions of even Fock states.}.

Several years later, different schemes for the generation of arbitrary
quantum states of light  -- including the superposition states
mentioned above -- have been proposed
\cite{Vogel1993,Parkins1993,Garraway1994,Law1996}.
All these schemes have in common that they are based on the
interaction of atoms with the single-mode light field of a cavity.
However, here we concentrate on the interaction dynamics of light and
single atoms in \emph{free space}, in particular in the dynamics of
spontaneous emission.
In other words, the squeezing in the fluorescence of a single atom
\cite{Walls1981,Lu1998} is a sensitive tool for studying the dynamics
of spontaneous emission in an unprecedented way.
Therefore, we sacrifice the possibility of generating arbitrary
quantum states and restrict the discussion to the most simple
superposition state, namely the one described in
Ref. \cite{Wodkiewicz1987}.
As we will outline below, this state can be generated by preparing a single
two-level atom in the corresponding superposition of its ground state and
its excited state and full solid angle collection of the spontaneously
emitted photon.  

The paper is organized as follows. 
In section \ref{th} we summarize the theoretical
model for obtaining the single photon squeezed state put forward
  in Ref. \cite{Wodkiewicz1987}. 
Section \ref{exp} is devoted to the discussion about the possibilities
for realization and detection of squeezing in superposition states in
a suitable free space experimental setup.

\section{Model}
\label{th}

\begin{figure}
\begin{center}
  \vbox{\vskip-2cm
  \scalebox{1.0}{\includegraphics{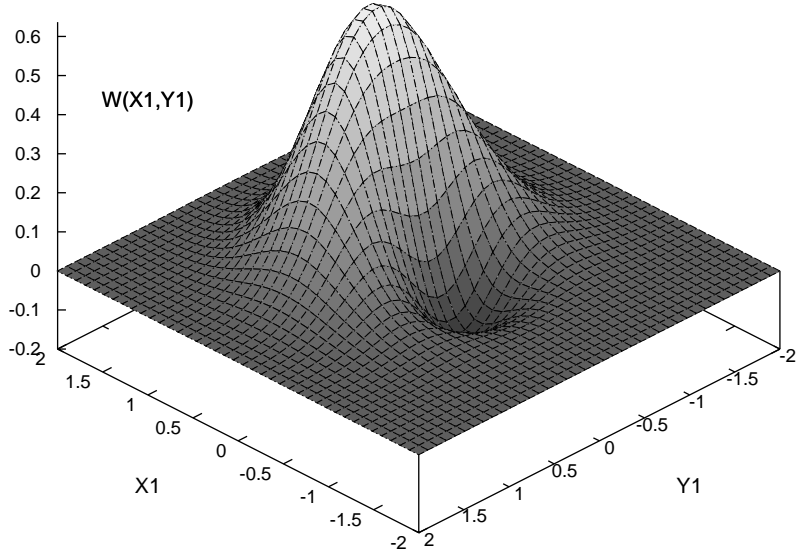}}
  \vskip-1cm}
\end{center}
\caption{Wigner function $W(X_1,X_2)$ evaluated for the one-photon
  superposition state in Eq. (\ref{sup}) for $\gamma = \sqrt{2/3}$ and
  $\beta = \sqrt{1/3}$.}
\label{Fig1}
\end{figure}

According to the squeezing transformation in Eq. (\ref{sq}), the
squeezed vacuum state is a superposition of
even Fock number states \cite{Yuen1976}
\begin{equation}
\hat{S}|0\rangle = e^{- \frac{1}{2}\ln\mathrm{ch \xi}} \exp^{-
  \frac{\mathrm{th}\xi}{2} {a^{\dagger}}^2} |0\rangle = e^{-
  \frac{1}{2}\ln\mathrm{ch \xi}} \Big\{ |0\rangle - \sqrt{2!}
\frac{\mathrm{th} \xi}{2}|2\rangle + \sqrt{4!} \frac{\mathrm{th}^2
  \xi}{4 \cdot 2!}  |4\rangle + ...\Big\},
\end{equation}
where the squeezing parameter $\xi$ was assumed to be real for
simplicity.  W\'odkiewicz et al. \cite{Wodkiewicz1987} discovered that
in one photon superposition states
\begin{equation}
|\varphi\rangle = \gamma|0\rangle + \beta|1\rangle
\label{sup}
\end{equation}
where $|\gamma|^2+|\beta|^2=1$, quadrature squeezing is present as
well for $|\beta|<1/\sqrt{2}$ and some special values of relative
phase $\phi$ between the probabilities amplitudes 
\begin{equation}
\label{eq:delta_superpos}
(\Delta X_1)^2(\phi = 0,\pi) = (\Delta X_2)^2(\pi/2, 3\pi/2) = 1/4 +
  |\beta|^2(|\beta|^2 - 1/2). 
\end{equation}
The values of $\phi = 0,\pi(\pi/2, 3\pi/2)$ correspond to the
amplitude (phase) quadrature squeezing respectively. The state
Eq. (\ref{sup}) is presented in terms of a Wigner function in Fig.
\ref{Fig1} for $\gamma = \sqrt{2/3}$ and $\beta = \sqrt{1/3}$.

The superposition given in Eq. (\ref{sup}) arises naturally during
interaction of a single quantized electromagnetic field mode initially
in the vacuum state with a two-level atom initially in a superposition
of excited and ground states. The system is described by the
Jaynes-Cummings Hamiltonian
\begin{equation}
H = 1/2 \hbar \omega_0 \sigma_z + \hbar \omega a^{\dagger}a + \hbar
\lambda (a^{\dagger} \sigma_- + \sigma_+ a)
\end{equation}
where $\omega_0$ is the atomic transition frequency, $\sigma_z$ is the
inversion operator, $\omega$ is the field mode frequency, $a$ and
$a^{\dagger}$ are field annihilation and creation operators, $\lambda$
is interaction strength, $\sigma_-$ and $\sigma_+$ are lowering and
rising atomic operators.
For the relation between the
Jaynes-Cummings single mode model and the Wigner-Weisskopf model with a
continuum of modes see Sec. \ref{exp} below.

Let us consider an atom prepared in a coherent superposition of
excited $|e\rangle$ and ground $|g\rangle$ states which interacts with
a vacuum field $|0\rangle$
\begin{equation}
|\Psi(t=0)\rangle = \cos(\theta/2) |e,0\rangle + e^{i \phi}
\sin(\theta/2) |g,0\rangle.
\end{equation}
The atom can spontaneously decay to its ground state and emit a
photon. 
In the Schr\"{o}dinger picture and on resonance the total wave
function of the system at time t reads 
\begin{eqnarray}
|\Psi(t)\rangle &=& \cos(\theta/2) \cos(\lambda t) e^{-i \omega t}
|e,0\rangle + e^{i \phi} \sin(\theta/2) |g,0\rangle \nonumber\\ & -& i
\cos(\theta/2) \sin(\lambda t) e^{-i \omega t} |g,1\rangle.
\end{eqnarray}
The uncertainties in quadrature
operators for this state measured by a homodyne detector are equal to
\begin{eqnarray}
\label{eq:dx1_a}
(\Delta X_1)^2 &=& 1/4 + \cos^2(\theta/2) \sin^2(\lambda t) \left(1/2
  - \sin^2\phi \sin^2(\theta/2) \right), \\
\label{eq:dx2_a}
(\Delta X_2)^2 &=& 1/4 + \cos^2(\theta/2) \sin^2(\lambda t) \left(1/2
  - \cos^2\phi \sin^2(\theta/2) \right).
\end{eqnarray}
The phase $\phi$ denotes the phase difference between the field and
the local oscillator in homodyne detection.
For $\theta = 2\pi/3,
4\pi/3$ and $\phi = \pi/2$ squeezing appears in the amplitude
quadrature
\begin{equation}
\label{eq:dx1}
(\Delta X_1)^2 = 1/4 - 1/16 \sin^2(\lambda t).
\end{equation}
Note that for time $t=\pi/2\lambda$ we obtain the desired
superposition Eq. (\ref{sup}). 
Furthermore, no squeezing arises for a purely excited
($\theta = 0$) or de-excited ($\theta = \pi$) atom.

As indicated by Eqs. (\ref{eq:dx1_a},\ref{eq:dx2_a}) the amount
of squeezing in the emitted photon superposition state is
dependent on the state in which the atom was prepared.
Thus, a precise measurement of the photonic state would give some
insight into the state -- possibly unknown a priori -- in which the
atom was before the emission process.

Moreover, in \cite{Wodkiewicz1987} a link between atomic dipole $D =
|g\rangle \langle e| = D_1 + i D_2$ squeezing and radiated field
squeezing has been established. If the initial atomic
state satisfies that $\langle[D^{\dagger},D]\rangle <0$ and the dipole
is squeezed $:(\Delta D_1)^2:<0$ the field will be squeezed as
well. Since in resonance fluorescence the normally ordered variance of
the electric field in the far field zone is related to the normally
ordered variance of the atomic dipole, it is possible to observe the
quadrature squeezing present in the one photon superposition state by
means of resonance fluorescence preparing the atomic state properly.

\section{Experimental prospects}
\label{exp}

The maximum amount of squeezing predicted by
Eqs. (\ref{eq:delta_superpos}) and (\ref{eq:dx1}) is 1.25 dB for
$|\beta|=1/2$.  This small reduction of quadrature fluctuations calls
for a sophisticated detection scheme.

First, the atom has to be prepared in the desired superposition state
by application of a suitable optical pulse.  Identifying $|\beta|$
with $|\cos(\theta/2)|$ determines the area of the excitation pulse
that prepares the atom in the superposition state that enables maximum
squeezing.  The state of the radiation field then has to be detected
transiently during the emission process via time resolved homodyne
measurements (e.g., \cite{Hansen2001,Wenger2004}).

Second, detection losses have to be minimized: The superposition state
has to be collected ideally over the complete solid angle of the
atomic emission, i.e., the full solid angle.
Recently, a setup based on a deep parabolic mirror has been
proposed that is capable of almost full solid angle coverage if the
atom is located in the focus of the mirror
\cite{Lindlein2007,Sondermann2007} 
(so far such a mirror does not yet exist with the required quality
being essentially aberration free, but the aberrations can be
corrected to a large extent by means of appropriate phase plates
\cite{Sondermann2009}).
E.g., if one monitors the
emission of a $\Delta\textrm{m}=0$ transition ($\pi$-polarization) by
an atom with its quantization axis oriented along the optical axis of
the parabolic mirror, current technology facilitates the collection of
94\% of the light emitted by the atom.
Thus, the full mode into which the atom emits can be detected.
We intend to use the
$^1\textrm{S}_0\leftrightarrow ^3\textrm{P}_1$ ground state transition
of $^{174}\textrm{Yb}^{2+}$.
It is planned to locate the ion in the focus of the parabolic
mirror by means of a needle-like ion trap.
First successful tests of such a trap geometry have been performed
recently \cite{Maiwald2009}. 
One might be concerned whether the parabolic mirror modifies the free
space modes.
We conjecture that a parabolic mirror with a focal length much larger
than the wavelength of the atomic transition -- as it is the case in
the planned setup -- does not change the
density of modes at its focal point (see Ref. \cite{Sondermann2007}
for a qualitative discussion and also Ref. \cite{Stobinska2009a}).

Furthermore, the mode of the local oscillator employed in the homodyne
measurement has to be matched to the atomic transition.  For known
emission characteristics, the mode after collection by the parabolic
mirror can be calculated in a straightforward fashion
\cite{Lindlein2007}.  In the case at hand, the mode profile of the
$\pi$-polarized emission after reflection off the mirror has a strong
overlap with radially polarized doughnut modes, which are easily
produced experimentally~\cite{Quabis2005}.

At this point one might wonder whether the single mode Jaynes-Cummings
model applied in the previous section is suitable for comparison with
the free space setting of our experimental setup.
To make the connection, note that in the best case the atom emits into
a single spherical dipole mode.
The problem can thus be treated by a one dimensional model
\cite{Stobinska2009}. 
One difference still remains: the free space atom emits into a
continuum of frequency modes.
We argue that such a superposition of pure states is itself a pure
state and effectively a single pulsed temporal mode.
One also might understand the mode operators $a$ and $a^{\dagger}$ as
operators for modes with a certain frequency distribution, as they are
employed in Ref.~\cite{Rohde2007}. 
For a photon generated by an atom in its
excited state that decays via interaction with all free space field
modes this distribution is Lorentzian.  The corresponding temporal
envelope of the photon is exponentially decaying as expected in free
space. 
In this sense the free space case should be comparable to the scenario
discussed by W\'odkiewicz et al.
The experiment will show whether this equivalence holds.

In other words, unlike in the Jaynes-Cummings model where the squeezed
superposition state occurs at every $t_{n}=(n+1/2)\pi/\lambda$ for
integer $n$, the photonic superposition state propagates away from the
atom after spontaneous decay.  The corresponding wave packet has
approximately the length of the atomic upper state lifetime.  In the
case of the $\textrm{Yb}^{2+}$ transition mentioned above the life
time is 230~ns \cite{Zhang2001}.  This is beneficial for time resolved
homodyne measurements, since the feasibility of this detection method
has been demonstrated with optical pulses of considerably smaller
duration \cite{Hansen2001}.

However, one has to be aware that if the squeezed superposition state
has a Lorentzian spectral distribution the local oscillator used in
homodyne detection should ideally have the same spectral shape.  We
intend to fulfill this requirement by cutting pulses out of continuous
wave (cw) laser beams by means of, e.g., acousto-optic modulators.
The cw laser has to be frequency stabilized to a line width much
smaller than the atomic transition line width.  Then, an exponentially
decaying pulse with its time constant matched to the atomic life time
and a length of a few life times spectrally matches the squeezed
superposition state almost perfectly.  Again, we emphasize that the
characteristics of the $^3\textrm{P}_1$ state of $\textrm{Yb}^{2+}$
appears to be advantageous for this task: The atomic line width of
approximately 700~kHz requires a cw laser that is stabilized to a line
width of about 1~kHz.  The latter number is well within the scope of
up-to-date technology.

\section{Conclusions}
More than two decades ago, K. W\'odkiewicz proposed to use one photon
superposition states for the generation of quadrature squeezed light
\cite{Wodkiewicz1987}. 
As outlined here, we argue that
such a realization is feasible in a setup based on a single ion in
free space when collecting the emitted radiation with a deep parabolic
mirror. 
This setup has initially been intended to facilitate efficient
absorption in free space \cite{Sondermann2007}.
But as elaborated in this paper, it can also be applied for the
measurement of the transient quantum properties of the field emitted
by a single atom. 
Hence, for the application in mind our setup poses a suitable
alternative to the schemes for quantum state generation based on
resonators \cite{Vogel1993,Parkins1993,Garraway1994,Law1996}, since
the latter hinder free space interaction between atom and photon by
design.

\section{Acknowledgments}

M. St. is partially supported by QAP European project.

\end{document}